\documentclass[cp1251]{jetp} 
\usepackage[cp1251]{inputenc}
\usepackage[english]{babel}
\usepackage{mathrsfs} 
\usepackage{cite}
\usepackage{caption}
\addto\captionsrussian{\renewcommand{\tablename}{Table}}
\addto\captionsrussian{\renewcommand{\refname}{References}}

\twocolumn

\let\figori=\fig \def\fig[#1]#2{\figori[#1]{*}}

\begin{document}

\title{FLUCTUATION MECHANISM OF SINGLE-ION ANISOTROPY  OF TOPOLOGICAL INSULATOR $\text{Mn}$$\text{Bi}_2$$\text{Te}_4$}

\rtitle{FLUCTUATION MECHANISM OF SINGLE-ION ANISOTROPY OF \dots \, $\text{Mn}$$\text{Bi}_2$$\text{Te}_4$}

\author{V.\,V.}{Val'kov}\email{vvv@iph.krasn.ru}
\affiliation{Kirensky Institute of Physics, Federal Research Center KSC SB RAS, Siberian Branch, Russian Academy of Sciences, Krasnoyarsk, Russia}

\author{ A.\,O.}{Zlotnikov}
\affiliation{Kirensky Institute of Physics, Federal Research Center KSC SB RAS, Siberian Branch, Russian Academy of Sciences, Krasnoyarsk, Russia}

\author{ A.}{Gamov}
\affiliation{Kirensky Institute of Physics, Federal Research Center KSC SB RAS, Siberian Branch, Russian Academy of Sciences, Krasnoyarsk, Russia}


\rauthor{VAL'KOV et al.}





\abstract{We demonstrate that charge fluctuations induced by electron hopping, combined with spin-orbit coupling, lift the sixfold degeneracy of the orbital singlet $^{6}S$ of Mn ions in the topological insulator MnBi$_2$Te$_4$, resulting in single-ion anisotropy. To solve the problem, a multiplet representation is introduced for the creation operators of atomic-state fermions in terms of the operators describing transitions between many-body wavefunctions. Using the operator form of perturbation theory up to the second order, we derive expressions for the populations $n_M$ of Mn ion states with spin projections $M$ of the $^{6}S$ term and determine the single ion anisotropy constants. The calculations reveal that the fluctuation mechanism ensures the possibility of  implementing the easy-axis anisotropy observed in MnBi$_2$Te$_4$. Notably, the range of anisotropy constants $D_2$ obtained by varying the model parameters includes the value $D_2 = -0.0095$ meV, required to reproduce the critical field of the spin-flop transition $H_{\text{sf}}$, known from the experiment . The proposed mechanism has a wide range of applicability for describing the anisotropy in compounds where the ground state of a magnetic ion in a weak crystal field is described by an orbital singlet.}

\maketitle

{\bf 1. INTRODUCTION}

Experimental data on the magnetic properties of
antiferromagnetic topological insulator
MnBi$_2$Te$_4$~
\cite{Otrokov_19, Yan_19, Hao_19, Li_20} show that in the low-temperature
area, the value of g-factor of divalent Mn ions is close
to 2, and the magnitude of spin $S =5/2$. Since the
nominal state of Mn$^{2+}$ ion in the marked compound
corresponds to the electronic configuration $3d^5$,
the given values of $g$-factor and spin indicate the
realization of the regime of weak crystal field~\cite{Goodenough_63, Eremenko_75}.
In this case, the intraatomic Coulomb interaction of
electrons prevails over the inter-site interaction.
Therefore, the ground state of the Mn$^{2+}$ ion
corresponds to the term ${}^{6}S$~\cite{Landau} with orbital moment $L=0$.

The equality to zero of the orbital moment for the
ground term of the manganese ion leads to the fact
that there should be no single-ion anisotropy (SIA)
in MnBi$_2$Te$_4$ in the linear in spin-orbit interaction (SOI)
parameter approximation.

Meanwhile, the magnetic properties of this compound in the antiferromagnetic phase suggest otherwise. In particular, the observation in the low temperature region of the spin-flop transition at increasing magnetic field applied along the trigonal axis indicates the presence of a rather strong uniaxial anisotropy of the “easy axis” type ~\cite{Otrokov_19, Hao_19, Yan_19-2, Lai_21}. This conclusion correlates also with the results of other experimental studies ~\cite{Li_24}.

The problem of microscopic description of SIA
for ions with orbital singlet has been discussed in
the theory of electron paramagnetic resonance ~\cite{Balkhauzen_1964, Alt_Koz_1972}. One scenario for the occurrence of SIA is related to the addition to the term ${^6}S$ of a term of high-energy configuration $3d^4 4s^1$, for which the orbital momentum is different from zero \cite{Balkhauzen_1964}. In this approach it is assumed that the influence of the environment of the ion in question can be reduced to the analysis of its energy spectrum in the external field. In reality, the ion, being in a crystal lattice, is able to give and receive electrons due to electron hopping. Due to such charge fluctuations, an orbital moment will be induced on the ion. When the SOI is taken into account, this will lead to a partial removal of the degeneracy of the orbital singlet.

In the paper~\cite{Li_19}, the density functional theory
(DFT) method was used to calculate the SIA parameter
at different values of the SOI intensity in Mn and Te
ions. The main conclusion was that the desired value
of the SIA constant appeared only in the case when
the SOI was taken into account simultaneously on
Te and Mn ions. At the same time, the mechanism
of SIA formation was associated with a change in the
electron density distribution on Mn ions, induced
by the modification of the charge distribution on Te
ions, arising from the consideration of the electron
SOI.

In the electronic ensemble condensed in the
form of many-electron states of Te and Mn ions,
the characteristic Coulomb interaction energy of
fermions significantly exceeds the average SOI
energy. Under these conditions, charge fluctuations
become more significant and lead to changes in
the electronic configurations of the marked ions,
initiating the mixing of states in which the orbital
moments are different from zero. Due to the SOI,
this causes a partial removal of the degeneracy of
the orbital singlet of the Mn$^{2+}$ ion and provides the
appearance of SIA. The described scenario of charge
transfer between Mn and Te ions forms the basis
of the fluctuation mechanism of SIA induction in
MnBi$_2$Te$_4$. The development of theoretical insights
in this direction forms the subject of this work.

The results of the study are summarized as follows. In Sect. 2 we describe the system under consideration and formulate the Hamiltonian in the atomic representation including the most essential fermion interactions. In Sect. 3 we introduce the multiplet representation for the Fermi operators via many-electron functions. Section 4 contains results of the operator form of perturbation theory in the multiplet representation. Section 5 is devoted to the derivation of an additional selection rule due to the consideration of the symmetry of the crystal
structure. Renormalized expressions for the filling
numbers of single-ion states with different values of
the spin projections of Mn ions are given in Sect. 6.
The obtained dependences of the SIA constant on the spin-orbit coupling parameters in tellurium and manganese ions are also discussed here. The results of the studies are summarized in the Conclusion.

{\bf 2. HAMILTONIAN OF THE ELECTRONIC SUBSYSTEM OF THE “TRILAYER” Te–Mn–Te}

In MnBi$_2$Te$_4$, a layer of Mn ions lies between two layers of Te ions. In each of these layers, the ions  are ordered in a triangular lattice. The Te ions in the layer above (below) the layer of Mn ions are displaced relative to the position of the Mn ions as shown in Fig.~1.

\begin{figure}[htb!]
    \begin{center}
        \includegraphics[width=0.52 \textwidth]{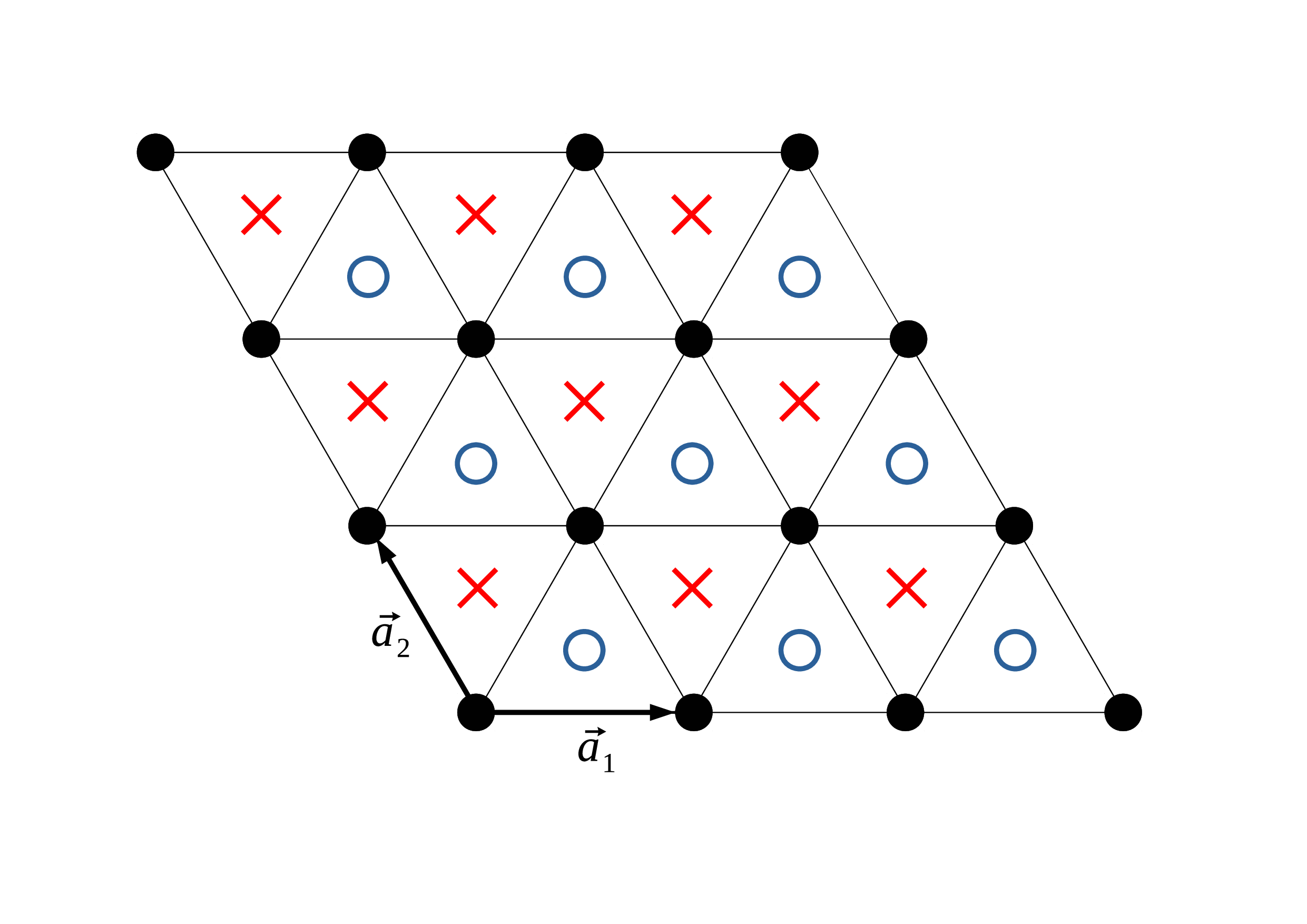}
        \caption{Fragment of the crystal structure of the Te–Mn–Te trilayer. The positions of Mn$^{2+}$ ions, indicated by dark circles,
        	form a triangular lattice in the middle plane of the trilayer. Above
        	and below them are Te$^{2-}$ ions, ordered according to the triangular lattice scheme. The projections of the centers of the upper Te$^{2-}$
        	ions onto the Mn$^{2+}$ arrangement plane are shown by red crosses.
        	Light circles indicate projections of the positions of Te$^{2-}$ ions
        	lying in the lower plane of the trilayer onto the midplane.}
    \end{center}
\end{figure}

We will solve the problem of the fluctuation
mechanism of SIA by taking into account the
electronic states of Mn and Te ions, which are part
of the Te–Mn–Te trilayer, as well as the basic
interactions. Let us write down the Hamiltonian of
the system in the form
\begin{eqnarray}
    \label{H_full}
    H=H_0 + V_{mix},
\end{eqnarray}
where the term
\begin{eqnarray}
    \label{H_0}
    H_0=\sum_{f}\sum_{m_{_d} \sigma}(\varepsilon_{_d} -\mu)d^{^\dagger}_{fm_{_d}\sigma}d_{fm_{_d}\sigma}+V^{d}_{c}+H^{d}_{SO}+ \nonumber\\
    \frac{1}{2}\sum_{f\delta}\sum_{m_{_p} \sigma}(\varepsilon_{_p} -\mu) p^{^\dagger} _{f+\delta, m_{_p}\sigma}p^{}_{f+\delta, m_{_p}\sigma}+H^{p}_{SO},
\end{eqnarray}
describes non-interacting electronic subsystems of
Mn and Te ions,  $\varepsilon_{_d}$ is energy of $d$ level, $\mu$ is chemical potential of the system, $d_{fm_{d}\sigma}$ $(d^{^\dagger}_{fm_{_d}\sigma})$ is annihilation (creation) operator of $d$ electron on Mn ion located in the site with number $f$, in the state with orbital momentum projection $m_{_d}$ $(m_{_d}=-2,-1,0,1,2)$
and spin projection $\sigma$ $(\sigma=\pm 1/2)$.

The Coulomb interaction operator of $d$ electrons
$V^{d}_{c}$ induces a splitting of the energy level of a given electronic configuration~\cite{Landau}. As a result, for the $3d^5$ configuration the lowest energy term corresponds to the orbital singlet $^{6}S$, while for the electronic configuration $3d^6$ the lowest energy term $^{5}D$ with orbital momentum $L=2$ and spin $S=2$ has the lowest energy. The states of these two terms will be taken into account in the following.

The SOI operator for manganese ions that have the state with the electronic configuration $3d^6$ (term $^5$D) is written in the following form
\begin{eqnarray}
    \label{Hd_SO}
    H^{d}_{SO}=-\sum_f \lambda_{_d}({\bf L}_f{\bf S}_{f}),
\end{eqnarray}
where ${\bf L}_f$ is the orbital momentum operator $(L_d=2)$, ${\bf S}_f$ is the spin operator $(S_d=2)$, for the above mentioned term of the Mn ion located at the site $f$.

The fourth summand in (\ref{H_0}) corresponds to the
account of electrons occupying $5p$ states of Te ions
located in sites with the number $f+\delta$, where
$\delta$ is the vector connecting the Mn ion to the nearest Te ion ($\delta$ takes 6 values as shown in Fig. 1). In this summand, $\varepsilon_{_p}$ is the energy of $5p$ orbital of the Te$^{2-}$ ion, $p_{_lm_{_p}\sigma}$ $(p^{^\dagger}_{_lm_{_p}\sigma})$ is the annihilation(creation) operator of the electron in the state with orbital momentum projection $m_{_p}=\pm1,0$ and spin projection $\sigma=\pm 1/2$ on the Te ion occupying the position with the number $l$.

The spin-orbit coupling operator for the subsystem of Te$^{1-}$ ions, which are in the electronic configuration $5p^5$, term ${^2}P$ with orbital momentum $L_p=1$ and spin $S_p=1/2$ is written similarly to the operator for manganese ions:
\begin{eqnarray}
    \label{Hp_SO}
    H^{p}_{SO}=-\sum_l \lambda_{_p}({\bf L}_l{\bf S}_{l}).
\end{eqnarray}
The parameters $\lambda_{_d}$ and $\lambda_{_p}$ determine the spin-orbit coupling intensities of Mn and Te ions for the marked terms, respectively.

In the considered mechanism of SIA formation, charge fluctuations due to hybridization of $p$ and $d$ electronic states, described by the operator, are
important.
\begin{eqnarray}
    \label{Vmix}
    V_{mix}=\sum_{f\delta}\sum_{m_{_p}m_{_d}\sigma}
    \left[ t_{m_{_d}m_{_p}}(\delta)d^{^\dagger}_{fm_{_d}
    \sigma}p_{f+\delta,m_{_p}\sigma} + \text{H.c.} \right],
\end{eqnarray}
in which the matrix element $t_{m_{_d}m_{_p}}(\delta)$ defines the amplitude of the intensity of the electron transition from $p$ state $|m_{_p}\sigma \rangle$ on the Te ion occupying the position with the number $f+\delta$, to $d$ state $|m_{_d}\sigma \rangle$ on the Mn ion located in the site with the number $f$.

 {\bf 3. HAMILTONIAN OF THE TRILAYER IN THE SPACE OF MULTIPLET STATES}

The transition of an electron from the Te ion to  the Mn ion changes the state of that ion: ${^6}S\rightarrow{^5}D$.
Since the states of the term with fixed values of the
orbital momentum projection and spin when the SOI
is included are not eigenstates of the Hamiltonian
describing isolated ions, in order to apply perturbation
theory to the degenerate level, it is necessary to go to
the well-known description in which the states of the
ion are reflected by the multiplet basis \cite{Landau}.

Such a description is based on the atomic
representation~\cite{Hubbard_65}, in which the Hamiltonian of the non-interacting between each other electronic subsystems of the Mn and Te ions, but taking into account spin-orbit coupling, takes a diagonal form:
\begin{eqnarray}
    \label{H0}
    H_{0}=\sum_{f} \left\lbrace   \sum_{M=-5/2}^{5/2}
    (E_{5}^{d}-5\mu)Z_{f}^{M;M} +\right. \nonumber\\
    \left.\sum_{J_{_d}}\sum_{M_{_J}}[E_{6}^{d}(J_{_d})-
    6\mu]Z_{f}^{J_{_d}M_{_J};J_{_d}M_{_J}}\right\rbrace   +\\
    \sum_{l}\left\lbrace (E_{6}^{p}-6\mu) X_{l}^{0;0} +
    \right. \nonumber\\
    \left.\sum_{J_{_p}}\sum_{m_{_J}} [E_{5}^{p}(J_{_p})-
    5\mu]X_{l}^{J_{_p}M_{_J};J_{_p}M_{_J}} \right\rbrace.\nonumber
\end{eqnarray}
In this expression, the Hubbard operator
\begin{eqnarray}
    Z_{f}^{M;M}=|f,M \rangle \langle\ f,M|\nonumber
\end{eqnarray}
is a projection operator on the orbital singlet
state ${}^{6}S$ of the Mn ion located at the site $f$, with the spin projection $S^z = M$ taking the values
$-5/2,-3/2,...,5/2$. The vector $|f,5/2 \rangle$ is defined by the expression
\begin{eqnarray}
    \label{Vecf52}
    |f,5/2 \rangle = d_{2\uparrow}^{\dagger}d_{1\uparrow}^{\dagger}
    d_{0\uparrow}^{\dagger}d_{\overline{1}\uparrow}^{\dagger}
    d_{\overline{2}\uparrow}^{\dagger}|f,0 \rangle
\end{eqnarray}
in which $|f,0 \rangle$ is the vacuum state, $\overline{m} = -m$.

Five other states with $M=3/2, 1/2,...,-5/2$ are
described by vectors obtained from \eqref{Vecf52} by successive application of the operator
\begin{eqnarray}
    S_{f}^- = \sum_{m=\overline{2}}^{2}d_{fm\downarrow}^\dagger d_{fm\uparrow}\nonumber
\end{eqnarray}
to the vector $|f,5/2 \rangle$ while taking into account the relation
\begin{eqnarray}
    S_{f}^-|f,M \rangle = \sqrt{(5/2+M)(5/2-M+1)}|f,M-1 \rangle.\nonumber
\end{eqnarray}

In (\ref{H0}) the energy of the sixfold degenerate term ${}^{6}S$ is denoted by $E_{5}^d$.

The second summand in the first curly brackets of
the operator $H_{0}$ (\ref{H0}) corresponds to the consideration
of the states of Mn ions in the configuration $3d^6$
with orbital momentum $L=2$ and spin $S=2$. It is
taken into account that due to the SOI the 25-fold
degenerate term ${}^{5}D$ is split into five multiplets . Each multiplet is characterized by the total momentum
$J_{_d}=4,3,2,1,0$ and energy
\begin{eqnarray}
    \label{Ed6}
    E_{6}^d(J_{_d}) = E_{6}({}^{5}D) - \lambda_{_d}\left[ J_{_d}(J_{_d}+1)/2-6\right],
\end{eqnarray}
where $E_{6}({}^{5}D)$ is the energy of the term ${}^{5}D$ of the configuration $3d^6$.

The diagonal Hubbard operators
\begin{eqnarray}
    Z_{f}^{J_{_d}M_{_J};J_{_d}M_{_J}}=|f,J_{_d}M_{_J} \rangle \langle\ f,J_{_d}M_{_J}|\nonumber
\end{eqnarray}
realize the projection of Hilbert space vectors onto
the multiplet state of the Mn ion, which has an
electronic configuration $3d^6$, an orbital momentum
$L=2$, a spin $S=2$, a total momentum $J_{_d}$ and a
projection of this momentum $M_{_J}$.

The summands $H_{0}$, standing in the second curly
brackets in (\ref{H0}), reflect the presence of Te ions having electron configurations $5p^6$ and $5p^5$. The energy of the singlet configuration is denoted by $E_{6}^p$. The operator
\begin{eqnarray}
    X_{l}^{0;0}=|l,p^6 \rangle \langle l,p^6|\nonumber
\end{eqnarray}
projects the Hilbert space vector onto the state of the
Te ion with a fully filled $5p$-shell:
\begin{eqnarray}
    |l,p^6 \rangle = p_{l1\uparrow}^{\dagger}p_{l1\downarrow}^{\dagger}p_{0\uparrow}^{\dagger}
    p_{0\downarrow}^{\dagger}p_{\overline{1}\uparrow}^{\dagger}p_{\overline{1}\downarrow}^{\dagger}|0 \rangle.\nonumber
\end{eqnarray}
The second summand, standing in the second
curly brackets in the Hamiltonian (\ref{H0}), accounts for the Te ion states appearing when one electron from
the $|5p^6 \rangle$ state moves to the Mn ion. The state $|l;J_{_p}M_{_J} \rangle$ arising at the site $l$ is characterized by the total momentum $J_{_p}$, which can take two values:  $J_{_p}=\frac{3}{2}$ and $J_{_p}=\frac{1}{2}$, as well as its projection $M_{_J}$.

The energy of a multiplet with a given value $J_{_p}$ is
defined by the expression
\begin{eqnarray}
    E_{5}^p(J_{_p}) = E_{5}^p - \lambda_{_p}\left[ J_{_p}(J_{_p}+1)/2-11/8\right].\nonumber
\end{eqnarray}

Let us introduce a multiplet representation for the
operators $d^{^\dagger}_{fm\sigma}$,
\begin{eqnarray}
    \label{dfm}
    d^{^\dagger}_{fm\sigma} = \sum_{JM}D_{m\sigma}(J;M)Z_{f}^{J,M+m+\sigma;M},
\end{eqnarray}
which allows us to consider the process of electron
appearance on the Mn$^{2+}$ ion as a coherent superposition of partial transitions of this ion from the state $|f;M \rangle$ to the state $|f;JM_{_J} \rangle$. Each of such transitions in the atomic representation is described by a non-diagonal Hubbard operator
\begin{eqnarray}
    Z_{f}^{JM_{_J};M}=|f;JM_{_J} \rangle \langle f;M|.\nonumber
\end{eqnarray}
The parameters of the representation (\ref{dfm}) can be
written as the product
\begin{eqnarray}
    \label{Dm}
    D_{m\sigma}(J;M) = \Gamma_{\sigma}(M)C_{J}(m,M+\sigma)
\end{eqnarray}
of the function
\begin{eqnarray}
    \label{Gamma}
    \Gamma_{\sigma}(M) = -2\sigma\sqrt{\frac{1}{2}-\frac{2M\sigma}{5}}
\end{eqnarray}
and the Clebsch-Gordan coefficient $C_{J}(m,M+\sigma)$~\cite{Landau}, which is a matrix element defining the decomposition
of the basis functions of the term ${}^{5}D$ by the basis
functions of the multiplets:
\begin{eqnarray}
    \label{fmM}
    |m,M+\sigma \rangle = \sum_{J}C_{J}(m,M+\sigma)|J,M+m+\sigma \rangle.
\end{eqnarray}

The annihilation operator of the electron on the Te
ion located at the site number $l$, from a state with orbital momentum projection $m=\pm1,0$ and spin projection $\sigma$ has a simpler form of multiplet representation:
\begin{eqnarray}
    \label{Plm}
    p_{lm\sigma} = \sum_{J_{_p}}K_{J_{_p}}(m,\sigma)X_{l}^{J_{_p},\overline{m}+\overline{\sigma};0}.
\end{eqnarray}
In this expression, the operator
\begin{eqnarray}
    \label{Xlm}
    X_{l}^{J_p,m+\sigma;0}=|l;J_p,m+\sigma \rangle \langle l,0|
\end{eqnarray}
describes the change of the state of the Te ion at the
site $l$ from the electronic configuration $5p^6$ (such
state corresponds to the Hilbert space vector $|l,0\rangle$) to the multiplet with the total momentum $J_{_p}=3/2,1/2$ and its projection $m+\sigma$, related to the electronic configuration $5p^5$.

The coefficients in \eqref{Plm} are defined by the expressions
\begin{eqnarray}
    \label{K}
    K_{3/2}(m\sigma)=\sqrt{\frac{2+2m\sigma}{3}},\nonumber\\
    K_{1/2}(m\sigma)=-2\sigma\sqrt{\frac{1-2m\sigma}{3}}.
\end{eqnarray}

The use of the obtained expressions allows us
to write down the operator $V_{mix}$ in the multiplet
representation necessary to apply the operator form
of perturbation theory for the degenerate level:
\begin{eqnarray}
    \label{VmixMP}
    V_{mix}=\sum_{f\delta}\sum_{m_{_p}m_{_d}\sigma}\sum_{J_{_d}J_{_p}M}\left\lbrace t_{m_{_p}m_{_d}}(\delta)\times\right.\nonumber\\
    \left.\times\Gamma_{\sigma}(m)C_{J_{_d}}(m_{_d},M+\sigma)K_{J_{_p}}(m_{_p},\sigma)\times\right.\\
    \left.\times Z_{f}^{J_{_d},M+m_{_d}+\sigma;M}X_{f+\delta}^{J_{_p},\overline{m}_{_p}+\overline{\sigma};0}+\text{H.c.}\right\rbrace.\nonumber
\end{eqnarray}

{\bf 4.  SPLITTING OF THE ORBITAL SINGLET ${}^{6}S$ OF THE ION Mn$^{2+}$}

The ground state Hamiltonian $H_{0}$ (\ref{H0}) corresponds to the nominal charge values of the Mn$^{2+}$ and Te$^{2-}$ ions.
In this case, the energy level $E_{5}^d$ of each manganese ion is sixfold degenerate. Consequently, the charge fluctuations can be accounted for only in the
framework of perturbation theory for the degenerate
level~\cite{Landau}. For this purpose it is convenient to use the operator form of perturbation theory~\cite{Bogolubov}.

Let us introduce a projection operator
\begin{eqnarray}
    \label{P}
    P=\left( \prod_{f}\sum_{M}Z_{f}^{MM} \right)\left(\prod_{l}X_{l}^{0;0} \right)
\end{eqnarray}
projecting the Hilbert space of the system under
study onto the subspace of nominal states of the
Hamiltonian $H_{0}$.

Considering $V_{mix}$ as a perturbation, we obtain
that in the second order of perturbation theory the effective Hamiltonian of the trilayer Te–Mn–Te
is written in the form of
\begin{eqnarray}
    \label{Heff}
    H_{eff}=\sum_{f}\sum_{M=-5/2}^{5/2} E_{5}^{d}Z_{f}^{M;M}+\sum_{l}E_{6}^{p}X_{l}^{0;0}+H_{eff}^{(2)},
\end{eqnarray}
where the second-order contribution, defined by the
expression
\begin{eqnarray}
    \label{Heff2}
    H_{eff}^{(2)}=-PV_{mix}\left(\frac{1}{H_{0}-E_0} \right) V_{mix}P,
\end{eqnarray}
leads to a partial removal of the degeneracy of the
term ${}^{6}S$, which, in accordance with the Wigner
Eckart theorem, can be described in terms of SIA.

To obtain the SIA Hamiltonian in explicit form,
we substitute into \eqref{Heff2} the hybridization operator in the multiplet representation \eqref{VmixMP} and consider the expression for the projection operator. Omitting simple intermediate calculations, we find the operator $H_{eff}^{(2)}$:
\begin{eqnarray}
    \label{Heff2_2}
    H_{eff}^{(2)}=-\sum_{f}\sum_{M^{'}M}V_{M^{'}M}^{(2)}Z_{f}^{M^{'}M},
\end{eqnarray}
where the matrix elements are of the form
\begin{eqnarray}
    \label{VMM2}
    V_{M^{'}M}^{(2)}=\sum_{m_{_p}m_{_d}}\sum_{m_{_p}^{'}m_{_d}^{'}} \left\lbrace  \sum_{\delta}t^{*}_{m_{_d}^{'}m_{_p}^{'}}(\delta)t_{m_{_d}m_{_p}}(\delta) \right\rbrace\times\nonumber\\
    \sum_{J_{_p}J_{_d}\sigma\sigma^{'}} \frac{\Gamma_{\sigma^{'}}(M^{'})\Gamma_{\sigma}(M) K_{J_{_p}}(m_{_p}^{'},\sigma^{'})K_{J_{_p}}(m_{_p},\sigma)}{\Delta_{dp}+E_{SO}^{d}(J_{_d}) +E_{SO}^{p}(J_{_p})}\times\nonumber\\
    C_{J_{_d}}(m_{_d}^{'},M^{'}+\sigma^{'})C_{J_{_d}}(m_{_d},M+\sigma)\times~~~~~~~~~\nonumber \\
    \Delta\left(m_{_p}^{'}+\sigma^{'}-m_{_p}-\sigma \right)\times~~~~~~~~~~~~~~~\\
    \Delta\left(m_{_d}^{'}+M^{'}+\sigma^{'}-m_{_d}-M-\sigma \right).~~~~~~~~~\nonumber
\end{eqnarray}
In this expression
\begin{eqnarray}
    \label{Deltdp}
    \Delta_{dp}=E({}^{5}D)-E_{5}^d+E_{5}^p-E_{6}^p\nonumber
\end{eqnarray}
defines the excitation energy at the transition of an
electron from the Te ion to the Mn ion without taking
into account the spin-orbit interaction, $\Delta( x-x')$ is the Kronecker symbol.

The values $E_{SO}^{d}(J_{_d})$ and $E_{SO}^{p}(J_{_p})$ denote the SOI energies for manganese and tellurium ions,
respectively:
\begin{eqnarray}
    \label{EdEpSO}
    E_{SO}^{d}(J_{_d}) = -\lambda_{_d}\left[ J_{_d}(J_{_d}+1)/2-6 \right], \nonumber\\
    E_{SO}^{p}(J_{_p}) = -\lambda_{_p}\left[ J_{_p}(J_{_p}+1)/2-11/8 \right].\nonumber
\end{eqnarray}

{\bf 5. TRIGONAL SYMMETRIES AND ADDITIONAL SELECTION RULES}

The previously noted triangular lattice type
ordering of tellurium and manganese ions, as well
as their mutual arrangement (see Fig. 1), lead to
important symmetry properties of the hopping integrals
$t_{m_{_d}m_{_p}}(\delta)$. To describe these properties, we introduce
the following notations. The three vectors connecting
the Mn ion with the three nearest Te ions located in
the plane above the Mn ion plane will be denoted
as $\delta_1$, $\delta_2$ and $\delta_3$. The increasing vector number corresponds to counterclockwise motion.

The other three vectors connecting the Mn ion
with the three nearest Te ions located under the Mn
ion plane will be denoted by the symbols $\delta_1^{'}$, $\delta_2^{'}$, $\delta_3^{'}$.
The correspondence between the vector number $\delta_i^{'}$ and its orientation is the same as for the vectors $\delta_i$.

The laws of transformations for spherical functions
at rotations about the third order axis lead to the
following
\begin{eqnarray}
    \label{tmdmp_del}
    t_{m_{_d}m_{_p}}(\delta_n) = \exp\left\lbrace -\frac{i2\pi}{3}(m_d-m_p)(n-1)\right\rbrace t_{m_{_d}m_{_p}}(\delta_1),\nonumber\\
    t_{m_{_d}m_{_p}}(\delta_n^{'}) = \exp\left\lbrace -\frac{i2\pi}{3}(m_d-m_p)(n-1)\right\rbrace t_{m_{_d}m_{_p}}(\delta_1^{'}), \nonumber \\
\end{eqnarray}
where $n=1,2,3$.

These relations give additional “selection rules” since
\begin{eqnarray}
    \label{tmdmp_sum}
    \left(\frac{2}{Z} \right) \sum_{\delta}t^{*}_{m_{_d}^{'}m_{_p}^{'}}(\delta)t_{m_{_d}m_{_p}}(\delta)=~~~~~~~~~~~~~~~~~~~~~~~~~~\nonumber\\
    \left[t^{*}_{m_{_d}^{'}m_{_p}^{'}}(\delta_1)t_{m_{_d}m_{_p}}(\delta_1) +t^{*}_{m_{_d}^{'}m_{_p}^{'}}(\delta^{'}_1)t_{m_{_d}m_{_p}}(\delta^{'}_1) \right] \times\nonumber\\
    \times\left[\Delta(m_{_d}-m_{_d}^{'}-m_{_p}+m_{_p}^{'}+3N_P)\right],~~~~~~~~~~~
\end{eqnarray}
where Z=6 is the number of nearest neighbors for
the manganese ion, $N_P=0, \pm1, \pm2,...$. The case
where $N_P \ne 0$, corresponds to the consideration
of transfer processes, the origin of which,
as usual, is a consequence of a discrete group
of transformations. In our consideration, these
processes do not contribute.

{\bf 6. HOPPING INTEGRALS}

It is known that first-principles calculations of the
electronic structure based on DFT methods allow us
to obtain the set of Bloch functions $\psi_{nk}({\bf r})$ and the energy spectrum of the system in K-space. With their help it is possible to define a set of Wannier functions $\phi_{m {\bf R}} ({\bf r})$, where \textbf{R} specifies the coordinates of the localization center of these functions. On the basis of such a basis, the hopping parameters are calculated as
averages of the DFT Hamiltonian over the $\phi_{m {\bf R}}$ basis
and the effective Hamiltonian in the strong coupling
approximation is determined. However, finding the
$\phi_{m {\bf R}}$ functions is difficult for a multiband electronic
structure due to the presence of overlapping bands,
and because there is uncertainty in the choice of the
phase factor for different $\psi_{nk}$ functions at a fixed $k$. As a result, the obtained Wannier functions may not be localized, generally speaking. To overcome this,
a numerical method has been proposed in which
the Wannier functions are found based on certain
combinations of functions $\psi_{nk}$, providing maximum
localization $\phi_{m {\bf R}}$ near ${\bf R}$ (see reviews~\cite{Marzari_12, Pizzi_20}).

For MnBi$_2$Te$_4$ the maximally localized Wannier
functions have been used in calculations of the electronic
structure of surface states and topological invariants~\cite{Otrokov_19, JLi_19, Xiao_22} (including the use of the topological materials
software package presented in~\cite{Wu_18}), as well as the Hall
conductivity~\cite{Gao_21, Guo_25} and the linear magnetoelectric
effect coefficient~\cite{Zhu_21}. Note that often the constructed
effective Hamiltonians for MnBi$_2$Te$_4$ take into account
only $p-p$ hoppings between Te–Te, Bi–Bi, and Te-Bi ions~\cite{Xiao_22, Guo_25} since these states exist near
the Fermi level and are important for describing the
properties noted above. However, to the best of our
knowledge, the calculation of the parameters of $p-d$
hoppings for MnBi$_2$Te$_4$, which play a determining role in
the formation of magnetic ordering in it (see also \cite{Li_20_1}),
has not been carried out on the basis of the described
approaches.

When finding the hopping integrals $~t_{m_{_d}m_{_p}}(\delta_1)$ and
$t_{m_{_d}m_{_p}}(\delta^{'}_1)$, let us take into account that the Bloch functions of the electronic states of the ions of the Te–Mn–Te trilayer are given by solutions of the
Schr\"{o}dinger equation with a Hamiltonian
\begin{eqnarray}
    \label{Bloch_equation}
    \hat{h}=-\frac{\hbar^2}{2m}\triangle+V_{eff}(r),
\end{eqnarray}
in which $m$ is the electron mass, $\triangle$ is the Laplace
operator, $V_{eff}(r)$ is the effective periodic potential
created by the ionic cores and the self-consistent field
arising from the Coulomb interaction of electrons.

Using the strong coupling method first applied
by Wilson to analyze the splitting of degenerate
electronic $p$ levels \cite{Wilson_1931}, we obtain that the integral of hoppings between the manganese ion and the
tellurium ion occupying with respect to the Mn
ion the position characterized by the vector $\delta$, is
determined by the expression
\begin{eqnarray}
    \label{defin_tpd}
t_{m_{_d}m_{_p}}(\delta)=\int\Phi^*_{d m_{_d}}(r)\hat{h}\Phi_{p m_{_p}}(r-\delta)dr,
\end{eqnarray}
where $\Phi^*_{d m_{_d}}(r)$ is the Wannier function corresponding
to $d$ state of the Mn ion located at the origin of the
coordinate system with orbital momentum projection $m_d$, $\Phi_{p m_{_p}}(r-\delta)$ is the Wannier function corresponding
to $p$ state of the Te ion located at the site defined by
the vector $\delta$, with orbital momentum projection $m_p$.

Since the periodic potential and Wannier functions
included in the definition of the hopping integral can
be found only approximately, we will further use
an approximation scheme. In accordance with it,
hydrogen-like functions for $3d$ and $5p$ states of
electrons participating in filling the electron shells
of manganese and tellurium ions, respectively, will
be used as functions $\Phi^*_{d m_{_d}}(r)$ and $\Phi_{p m_{_p}}(r-\delta)$. In
this case, the mean-field potential created by the
ionic shells and the Coulomb repulsion potential of
electrons is used as $V_{eff}(r)$.

It should be emphasized that the values of
$t_{m_{_d}m_{_p}}(\delta)$ obtained by this approach can correspond
to the true values only with an accuracy up to a
coefficient of the order of one. It is significant,
however, that when this coefficient is changed, the
resulting range of values of the single-ion anisotropy
parameter corresponds to the value known from
experimental data on the spin-flop transition. It is
also important that the applied method of calculation fully preserves the above mentioned symmetry
properties of the integrals $t_{m_{_d}m_{_p}}(\delta)$.

\begin{table}[h!]
    \centering
    \begin{tabular}{|c|c|c|c|}
        \hline
        $m_d \backslash m_p$ & 1 & 0 & -1 \\ [1ex] \hline
        2 & -0.31 + 0.54$i$ &  0.05 - 0.03$i$ & -0.06  \\ \hline
        1 &  - 0.37$i$ & 0.20 - 0.35$i$ & 0.07 - 0.04$i$  \\ \hline
        0 & -0.13 - 0.22$i$ & 0.53$i$ & -0.13 + 0.22$i$  \\ \hline
        -1 &  -0.07 - 0.04$i$ & 0.20 + 0.35$i$ & -0.37$i$ \\ \hline
        -2 & -0.06 & -0.05 - 0.03$i$ & -0.31 - 0.54$i$ \\
        \hline
    \end{tabular}
    \caption{Values of the hopping integrals $t_{m_dm_p}(\delta_1)$ in
    	electronvolts between the Mn ion and the nearest Te ion
    	in the upper plane of the Te–Mn–Te trilayer occupying
    	the position characterized by the vector $\delta_1$.}
\end{table}

Table 1 summarizes the hopping integrals for different
values of $m_d$ and $m_p$, if $\delta=\delta_1$. In a Cartesian coordinate
system with the axis $x$, directed along the vector $\bf{a_1}$, (see
Fig. 1) and the axis $z$, oriented perpendicular to the
Mn ion layer, this vector is defined by the expression
$\delta_1=(a/2, ~-a/(2\sqrt 3), ~h)$, where $a=4.28$~{\AA} is the
distance between the nearest Mn ions, $h=1.66$~{\AA} is
the distance between the Mn and Te ion planes.
The values of the hopping integrals for the vector
$\delta_1^{\prime} =(a/2, ~a/(2\sqrt 3), ~-h)$ are summarized in Table 2.
The values of $t_{m_d,m_p}(\delta)$ for other $\delta$ satisfy the conditions noted above (\ref{tmdmp_del}).

\begin{table}[h!]
    \centering
    \begin{tabular}{|c|c|c|c|}
        \hline
        $m_d \backslash m_p$ & 1 & 0 & -1 \\ [1ex] \hline
        2 & 0.31 + 0.54$i$ &  0.05 + 0.03$i$ & 0.06  \\ \hline
        1 & 0.37$i$ & -0.20 - 0.35$i$ & 0.07 + 0.04$i$  \\ \hline
        0 & 0.13 - 0.22$i$ & -0.53$i$ & 0.13 + 0.22$i$  \\ \hline
        -1 & -0.07 + 0.04$i$ & -0.20 + 0.35$i$ & 0.37$i$ \\ \hline
        -2 & 0.06 & -0.05 + 0.03$i$ & 0.31 - 0.54$i$ \\
        \hline
    \end{tabular}
    \caption{Values of hopping integrals $t_{m_d   m_p}(\delta_1^{\prime})$ in electronvolts between the Mn ion and the nearest Te ion
    in the lower plane of the Te–Mn–Te trilayer occupying
    the position characterized by the vector $\delta_1'$.}
\end{table}

{\bf 7. UNIAXIAL ANISOTROPY CONSTANT AND RENORMALIZATION OF THE ENERGY THERM ${}^{6}S$}

When the additional selection rule (\ref{tmdmp_sum}) is taken into account, the operator $H_{eff}^{(2)}$ is diagonalized by
indices $M$ and $M^{'}$:
\begin{eqnarray}
    \label{Heff3}
    H_{eff}^{(2)}=\sum_{f}\sum_{M}E_{M}^{(2)}Z_{f}^{MM},\nonumber
\end{eqnarray}
where
\begin{eqnarray}
    \label{EM2} E_{M}^{(2)}=-\left(\frac{Z}{2}\right)\sum_{\sigma\sigma'}\Gamma_{\sigma^{'}}(M)\Gamma_{\sigma}(M)
    \sum_{m_{_p}m_{_d}}T_{m_{_d}m_{_p}}^{\sigma,\sigma'}\times \nonumber\\
    \sum_{J_{_p}J_{_d}}\left\{\frac{K_{J_{_p}}(m_{_p}+\sigma-\sigma',\sigma') K_{J_{_p}}(m_{_p},\sigma)}{\Delta_{dp}+E_{SO}^{d}(J_{_d})+E_{SO}^{p}(J_{_p})}\times\right.\nonumber\\
    \left. C_{J_{_d}}(m_{_d}+\sigma-\sigma',M+\sigma')C_{J_{_d}}(m_{_d},M+\sigma)\right\}.
\end{eqnarray}
In this expression, $T_{m_{_d}m_{_p}}^{\sigma,\sigma'}$ is a combination of products of hopping integrals
\begin{eqnarray}
    T_{m_{_d}m_{_p}}^{\sigma,\sigma'}=t^*_{m_{_d}+\sigma-\sigma',m_{_p}+\sigma-\sigma'}(\delta_1)
    t_{m_{_d},m_{_p}}(\delta_1)+\nonumber\\
    t^*_{m_{_d}+\sigma-\sigma',m_{_p}+\sigma-\sigma'}({\delta}'_1)
    t_{m_{_d},m_{_p}}({\delta}'_1).~~~~
\end{eqnarray}
From the symmetry properties of the Clebsch–Gordan coefficients, the function $\Gamma_{\sigma}(M)$ and the hopping integrals $t_{m_{_d}m_{_p}}$, it follows that $E^{(2)}_M=E^{(2)}_{-M}$.
This corresponds to the character of the term splitting with $S=5/2$ in the uniaxial anisotropy field.

Taking into account the sum rule
\begin{eqnarray}
    \sum_{M}Z_{f}^{MM}=1 \nonumber
\end{eqnarray}
we obtain that the operator expression
\begin{eqnarray}
    R_{f}=\sum_{M}E^{(2)}_{M}Z_{f}^{MM} \nonumber
\end{eqnarray}
can be represented in the form
\begin{eqnarray}
    \label{Rf}
    R_{f}=\Delta E^{(2)}\hat{I}_{6f}+B^0_{2} O^0_{2f}+B^0_{4}O^0_{4f},
\end{eqnarray}
where
\begin{eqnarray}
    O^0_{2f}=3\left( S_{f}^z \right)^2 - S(S+1),~~~~~~~~~~~~~~~~\\
    O^0_{4f}=35\left( S_{f}^z \right)^4-\left[ 30S(S+1)-25\right]\left( S_{f}^z \right)^2 +\nonumber\\
    +3S^2(S+1)^2-6S(S+1)  ~~~~~~~~~\nonumber
\end{eqnarray}
are the Stevens operators \cite{Stewens1952}, used in the
description of anisotropy on the basis of the
equivalent Hamiltonian \cite{Zvezdin1985}. By means of $\hat{I}_{6f}$, we
denote a unit matrix of size $6\times6$, related to the site
number $f$.

The fluctuation shift $\Delta E^{(2)}$ defines the
renormalization of the energy of the term $^{6}S$, arising
in the second order of perturbation theory and
leading to a lowering of the energy of this term. Its
magnitude and the values of the SIA parameters $B^0_{2}$
and $B^0_{4}$ are related to the energies $E^{(2)}_M$ by linear equations:
\begin{eqnarray}
\label{B2m}
\Delta E^{(2)}=(E^{(2)}_{1/2}+E^{(2)}_{3/2}+E^{(2)}_{5/2})/3, \nonumber\\
B^0_{2}=(5E^{(2)}_{5/2}-4E^{(2)}_{1/2}-E^{(2)}_{3/2})/84,  \\
B^0_{4}=(E^{(2)}_{5/2}+2E^{(2)}_{1/2}-3E^{(2)}_{3/2})/840.
\nonumber
\end{eqnarray}
Using the data of Tables 1 and 2, as well as
the expressions (\ref{EM2}) and (\ref{B2m}) allows us to find the dependences of the value $D_2$ (this parameter
corresponds to writing the anisotropy energy operator
in the form $H_A=D_2(S^z)^2$, with the following
relation $D_2=3B_2^0$) on spin-orbit coupling constants
$\lambda_d$ and $\lambda_p$. The value of $\Delta_{dp}=10$~eV.
\begin{figure}[htb!]
    \begin{center}
        \includegraphics[width=0.52 \textwidth]{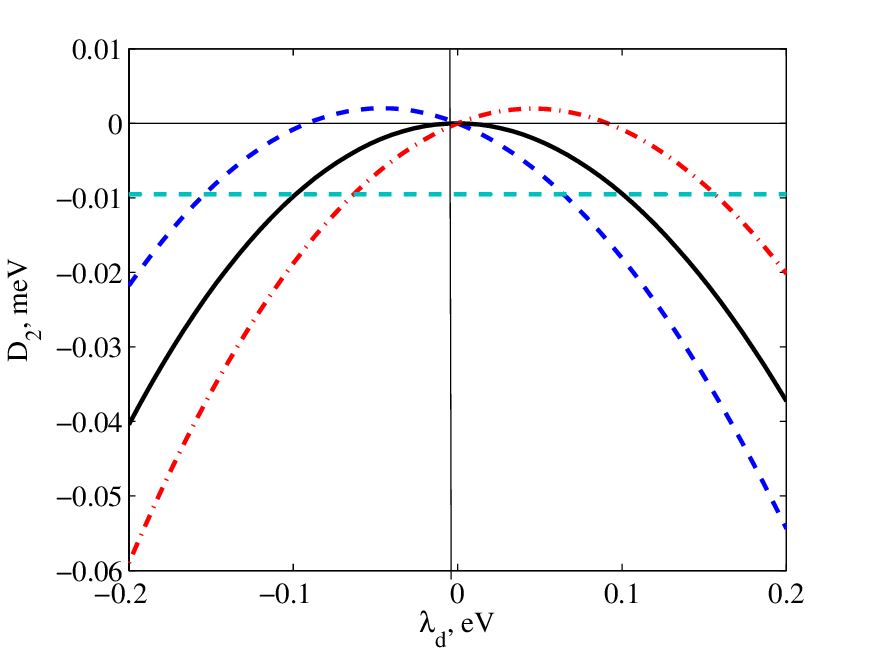}
        \caption{Dependences of the anisotropy parameter $D_2$ on $\lambda_d $ at
        	three values of $\lambda_p$. The black solid line is plotted for $\lambda_p = 0.$ The
        	red dashed line corresponds to the value $\lambda_p = 0.02$~eV. The blue dashed
        	line is plotted for $\lambda_p = -0.02$~eV. The horizontal dashed line
        	indicates the value of $D_2 = -0.0095 meV$ corresponding to the
        	value of the SIA parameter of MnBi$_2$Te$_4$, determined from spin-flop
        	field data.}
    \end{center}
\end{figure}

The results of the calculations are presented
graphically in Figs. 2 and 3. The first of them shows
the dependences of the value of $D_2$ on the spin
orbit coupling constant $\lambda_d$ at three values of the SOI
parameter $\lambda_p=-0.02$~eV,~$\lambda_p=0$,~$\lambda_p=0.02$~eV.

If $\lambda_p=0$, only anisotropy of the “easy-axis” type is realized at all $\lambda_d$, when
$D_2 < 0 $. The dependence of $D_2(\lambda_d)$ for this case
is shown in Fig. 2 by the black line. The dashed
horizontal line in this figure corresponds to the value
$D_2 =D^{(exp)}_2 =-0.0095$ meV, which is obtained from
the processing of experimental data on the spin-flop transition in MnBi$_2$Te$_4$ \cite{Otrokov_19, Hao_19, Yan_19-2, Lai_21, Li_24, Valkov_24}. The
two points of intersection of this line and the black
curve indicate that the value $D_2 < 0$ required for the
experiment takes place both at $\lambda_d > 0$ and at $\lambda_d < 0$.
Note that in this case the marked intersections occur
at a sufficiently large value of the absolute value of $\lambda_d$.

When including even weak SOI for the excited
term of the Te ion ($|\lambda_p| = 0.02$~eV), there is a
qualitative change in the dependence $D_2(\lambda_d)$. First
of all, the symmetry with respect to the change of sign $\lambda_d $ disappears, since there is a shift of the curve
relative to the origin of coordinates.

At positive $\lambda_p$ the graph shifts upward and to
the right (red dashed line in Fig. 2). This leads to a
significant modification of the phase diagram of the
system, since in the right vicinity of the point $\lambda_d = 0$
there appears an interval of change $\lambda_d $, within which
anisotropy of the “easy-plane” type is realized. At
that, “easy-axis” anisotropy takes place only
starting from some value $\lambda_d > 0$.

In the region of negative $\lambda_d $, “easy-axis” anisotropy is realized at all $\lambda_d$. For practical
purposes it is essential that in this case the value of
$D_2 =D^{(\exp)}_2$ is reached at a smaller value of $ |\lambda_d| $
compared to the case of $\lambda_p = 0$.

At negative $\lambda_p $ (blue dashed line in Fig. 2) the
graph shifts upward and to the left. It is easy to see
that in this case there appears a small interval of
negative values $\lambda_d $, within which “easy-plane” anisotropy is realized . Beyond this interval,
anisotropy of the “easy-axis” type takes place.

As in the first case, the following regularity
takes place: if the product $\lambda_d $ $\cdot$ $\lambda_p $ $< 0 $, then the value $D_2 =D^{(\exp)}_2 $ is reached at smaller $ |\lambda_d| $. If $\lambda_d$ $\cdot$ $\lambda_p$ $> 0$ ,
then to obtain the required value of $D_2 =D^{(exp)}_2$, it
is necessary to take larger $|\lambda_d| $.

The modification of the dependences $D_2(\lambda_d)$
at significant intensity of SOI on the excited states
of Te ions is shown in Fig.3. The main conclusion is that for values of $|\lambda_p|\simeq 0.1$~eV the “easy-axis”
anisotropy is realized only when the condition
$\lambda_d$ $\cdot$ $\lambda_p $ $< 0 $ is fulfilled, since the second branch of the
parabola crosses the line $D_2=0$ at unrealistically
large constants $|\lambda_d|$.

\begin{figure}[htb!]
    \begin{center}
        \includegraphics[width=0.52 \textwidth]{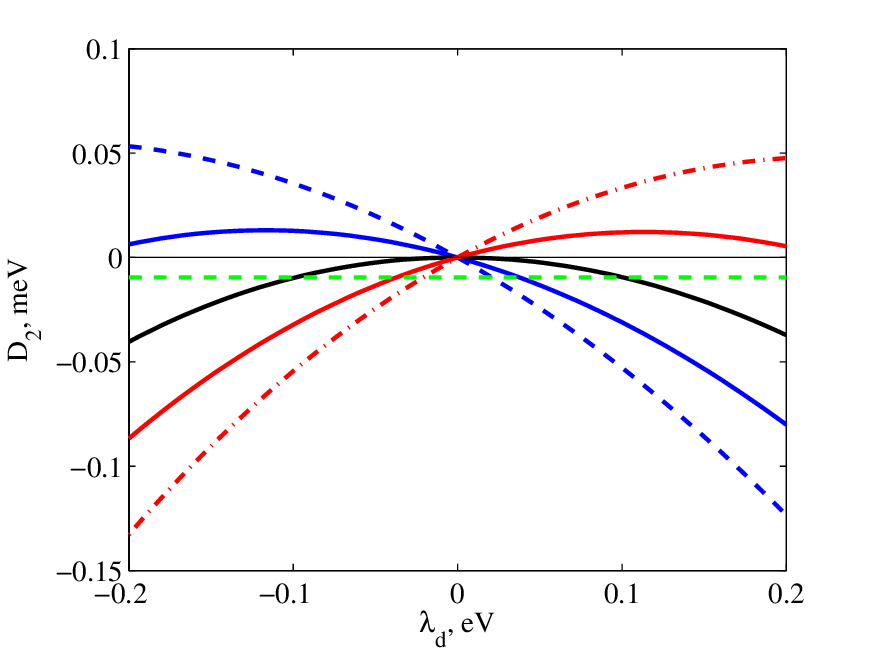}
        \caption{Dependencies of $D_2(\lambda_d)$ at five values of the $\lambda_p$ parameter:
        	$\lambda_p =-0.1$~eV (red dashed line),  $\lambda_p =-0.05$~eV (red solid line),
        	$\lambda_p =0.1$~eV (blue solid line), up to $\lambda_p =0.1$~eV (blue dashed
        	line). The black solid line corresponds to the value of $\lambda_p = 0$. The
        	horizontal dashed line has the same meaning as in Fig.~2.}
    \end{center}
\end{figure}

The presented results correlate with the
conclusions obtained on the basis of numerical DFT
calculations \cite{Li_19}. In this paper it was noted that the agreement with the experimental data occurs when
both $\lambda_p$ and $\lambda_d$ are simultaneously different from zero. In our case this conclusion has to be made
because for $\lambda_p = 0$ the value of $\lambda_d$, which yields $D_2 = -0.0095$ meV, is too large. At the same time,
for $\lambda_p\neq 0$ the necessary value of the anisotropy
parameter is achieved at much smaller $\lambda_d$.

Parameter $B_4^0=0$, since in second-order
perturbation theory, irreducible summands of degree
four do not arise.

The fluctuation shift ~$\Delta E^{(2)}$, corresponding to the reduction of the single-ion energy in the second
order perturbation theory, is of the order of ~$\sim-1$~eV at the chosen values of the system parameters.

{\bf 8. PARAMETER $D_2$ AND SHIFT $\Delta E^{(2)}$
	AT  $\Delta_{dp}\gg E_{so}$}

At real values of $\Delta_{dp},~ \lambda_d,~ \lambda_p $ the ratio

\begin{eqnarray}
    \label{par_epsilon}
    |E_{SO}^{d}(J_{_d})+E_{SO}^{p}(J_{_p})|/\Delta_{dp} \equiv \epsilon \ll 1
\end{eqnarray}
forms a parameter of smallness. Therefore,
the denominator of the expression (\ref{EM2}) can be
decomposed into a series by $\epsilon$. The summation over $\sigma,~\sigma',~ J_p$ and $J_d $ can then be done explicitly.

Taking into account the orthonormalization
relations of the Clebsch-Gordon coefficients,
their symmetry properties, as well as the equality
$|t_{m_d m_p}(\delta)|=|t_{m_d m_p}(\delta')|$, the validity of which is easy to establish on the basis of the data of Tables 1 and 2, we obtain that the dependences of the values of $ E^{(2)}_M $ on the spin-orbit coupling constants with quadratic accuracy on the parameter $\epsilon$ are determined by the expressions
\begin{eqnarray}
    \label{E2M}
    E^{(2)}_{5/2} =-A-B\lambda_p^2+C\lambda_p\lambda_d-F_{5/2}\lambda_d^2,~~~ \nonumber\\
    E^{(2)}_{3/2} =-A-B\lambda_p^2+C\lambda_p\lambda_d/5-F_{3/2}\lambda_d^2, \\
    E^{(2)}_{1/2} =-A-B\lambda_p^2-C\lambda_p\lambda_d/5-F_{1/2}\lambda_d^2\nonumber,
\end{eqnarray}
where
\begin{eqnarray}
    \label{ABC}
    A=Z\sum_{m_{_d}m_{_p}}\frac{|t_{m_{_d}m_{_p}}|^2}{\Delta_{dp}},~~~~ B=\frac{A}{2\Delta^2_{dp}},~~~~~~~~~~~\\
    C=4Z\left(\frac{|t_{1,1}|^2-|t_{-1,1}|^2+2|t_{2,1}|^2-2|t_{-2,1}|^2}{\Delta^3_{dp}}\right),
\end{eqnarray}
and the $M$-dependent coefficients of $F_M$ are represented as follows
\begin{eqnarray}
    \label{FM}
    F_{5/2}=\frac{6Z}{\Delta^3_{dp}}\sum_{m_{_p}}\left(6|t_{2,m_{_p}}|^2+
    3|t_{1,m_{_p}}|^2+|t_{0,m_{_p}}|^2\right),\nonumber\\
    F_{3/2}=\frac{6Z}{5\Delta^3_{dp}}\sum_{m_{_p}}(18|t_{2,m_{_p}}|^2+
    21|t_{1,m_{_p}}|^2+11|t_{0,m_{_p}}|^2),\nonumber\\
    F_{1/2}=\frac{12Z}{5\Delta^3_{dp}}\sum_{m_{_p}}(6|t_{2,m_{_p}}|^2+
    12|t_{1,m_{_p}}|^2+7|t_{0,m_{_p}}|^2)\nonumber.
\end{eqnarray}

Using the relation (\ref{B2m}), as well as the obtained
expressions for $F_M$ allows us to obtain a simple
formula for the anisotropy parameter
\begin{eqnarray}
    \label{D2}
    D_2~=~-F \lambda_d^2~+~\frac{C}{5}\lambda_p \lambda_d,
\end{eqnarray}
in which
\begin{eqnarray}
    \label{F}
    F ~=~\frac{9Z}{5\Delta^3_{dp}}\sum_{m_{_p}}\left(2|t_{2,m_{_p}}|^2-
    |t_{1,m_{_p}}|^2-|t_{0,m_{_p}}|^2\right).
\end{eqnarray}

Since the coefficients at the summands $\sim \lambda_p^2$ in (\ref{E2M}) are the same, according to (\ref{B2m}), $D_2$ is independent of $\sim \lambda_p^2$. However, such summands give a contribution to the renormalization of the energy of the term ${}^6S$, which can be written as
\begin{eqnarray}
    \label{DeltaE2}
    \Delta E^{(2)}=-A\left(1+\frac{1}{2}\frac{\lambda_p^2}{\Delta^2_{dp}}
+12\frac{\lambda_d^2}{\Delta^2_{dp}}\right)
+\frac{C}{3}\lambda_p\lambda_d.
\end{eqnarray}

Tables 1 and 2 show that for the considered case
\begin{eqnarray}
    \label{reltpd}
    |t_{1,1}| \gg |t_{-1,1}|,~~~|t_{2,1}| \gg |t_{-2,1}|.
\end{eqnarray}
Therefore, the value
\begin{eqnarray}
    \label{C}
     C\simeq 4Z\left(\frac{|t_{1,1}|^2+2|t_{2,1}|^2}{\Delta^3_{dp}}\right) \simeq
     \left(\frac{4Z}{\Delta^3_{dp}}\right)0.91
\end{eqnarray}
is obviously positive.

On the other hand, for the value $F$, the
contribution of the summands with negative signs in
(\ref{F}) is commensurate with the contribution of the
positive summands. As a result, compensation occurs
so that a small multiplier appears in the expression
for $F$:
\begin{eqnarray}
    \label{F_appr}
    F ~\simeq~\frac{9Z}{5\Delta^3_{dp}}\cdot 0.089.
\end{eqnarray}

From a comparison of the expressions (\ref{C}) and (\ref{F_appr}), the inequality $C \gg F$ which explains the correlation of the results obtained with the results of calculations of the parameter $D_2$ by the DFT method \cite{Li_19}. It was noted in \cite{Li_19} that $\lambda_d$ and $\lambda_p$ must be different from zero to obtain the required value of the uniaxial anisotropy parameter.

In our case at $\lambda_p=0$, the uniaxial anisotropy
parameter is determined by the expression $D_2=-F \lambda_d^2$. Since the value of $F$ is relatively small,
the necessary value of $D_2$ arises only at large values
of $\lambda_d$.

The situation is different at $\lambda_p \neq 0$, when the
second summand in (\ref{D2}) is included. Since $C \gg F$,
the required value of the parameter $D_2$ even at
relatively small $\lambda_p$ is achieved at much smaller values of $\lambda_d$.

The analysis of analytical expressions is fully
consistent with the results of \cite{Li_19}, as well as with
the above graphs defining the dependences of $D_2$
on $\lambda_p$ and $\lambda_d$, which were obtained on the basis of numerical calculations without using the above
mentioned series expansion. Note also that the
obtained analytical formulas allowed us to trace in
detail the influence of spin-orbit coupling parameters
on the region of realization of anisotropy of the “easy-axis” and “easy-plane” types.

It should be emphasized that the values of $F$
and $C$ in (\ref{D2}) depend significantly on the hopping
integrals, which in turn change with the change of
the geometrical arrangement of ions with respect to
each other.

The obtained results suggest the possibility of
a wider application of the proposed fluctuation
mechanism of SIA formation in insulators with
magnetically active ions in the orbital singlet state.

{\bf CONCLUSION}

The fluctuation mechanism of single-ion
anisotropy in antiferromagnetic topological insulator MnBi$_2$Te$_4$ proposed in this
paper is based on the use of the following statements.

1. The states with nominal valence of Mn$^{2+}$ and
Te$^{2-}$ ions correspond to electronic configurations $3d^5$ and $5p^6$ with terms ${}^{6}S$ and ${}^{0}S$, respectively. In this case, the term ${}^{6}S$ is sixfold degenerate.

2. Electron hopping between Te and Mn ions
initiates charge fluctuations. As a result, there is
an admixture to the states of the basic terms of the
marked ions of the states of excited terms with non
zero orbital moments.

3. To account for these processes, the operator
form of perturbation theory is applied with the
involvement of the atomic representation, which
allows us to write down the multiplet states in
diagonal form. In the second order of perturbation
theory the effective Hamiltonian is obtained, in
which the partial removal of degeneracy of the term
${}^{6}S$ corresponds to the appearance of single-ion
anisotropy.

4. Simple analytical expressions describing the
dependence of the single-ion anisotropy constant on
the parameters of spin-orbit interactions for Te and
Mn ions are obtained using a smallness parameter.

5. The interval of values of the anisotropy constant
arising from varying the initial model parameters is
found to be important for applications and includes
that value $D_2$, which is required for the interpretation
of experimental data on the spin-flop transition in MnBi$_2$Te$_4$.

6. The fluctuation mechanism of SIA considered in
this work is not fundamentally limited to application
only to MnBi$_2$Te$_4$, but can be obviously adapted to
other compounds containing magnetically active ions
with orbital singlet as the main term.

{\bf FUNDING}

This study was supported by the grant of the
Russian Science Foundation No. 23-22-10021,
\emph{https://rscf.ru/project/23-22-10021/} and the
Krasnoyarsk Regional Science Foundation

\emph{Translated by Nauka Publishing House.}

\end{document}